\documentclass[12pt]{article}
 \usepackage[ruled,vlined,linesnumbered]{algorithm2e}
\usepackage{booktabs,amsfonts,color}
\usepackage{graphicx}
\usepackage{amsmath,bm}
\usepackage{comment}
\usepackage{multirow}
\usepackage{setspace}
\usepackage{float}
\usepackage{color}
\usepackage{indentfirst}
\usepackage{epstopdf}
\usepackage{epsfig}
 
 \usepackage{natbib}
 
\date{}
\begin{document}
 \def\spacingset#1{\renewcommand{\baselinestretch}%
{#1}\small\normalsize} \spacingset{1}

 \title{ Exploring the space-time pattern of log-transformed  infectious count of COVID-19: a clustering-segmented
	autoregressive sigmoid model}

\author{Xiaoping Shi\thanks{Department of Mathematics and Statistics, Thompson Rivers University,  email: xshi@tru.ca} 
\hspace{.3cm} 
Meiqian  Chen and Yucheng Dong\thanks{Center for Network Big Data and Decision-Making, Business School, Sichuan University,   email: ycdong@scu.edu.cn}}
  \date{}
  \maketitle
 
\begin{abstract} 
At the end of April 20, 2020, there were only a few new COVID-19 cases remaining in China, whereas the rest of the world had shown increases in the number of new cases. It is of extreme importance to develop an efficient statistical model of COVID-19 spread, which could help in the global fight against the virus. We propose a clustering-segmented autoregressive sigmoid (CSAS) model to explore the space-time pattern of the log-transformed infectious count. Four key characteristics are included in this CSAS model, including unknown clusters, change points, stretched S-curves, and autoregressive terms, in order to understand how this outbreak is spreading in time and in space, to understand how the spread is affected by epidemic control strategies, and to apply the model to updated data from an extended period of time. We propose a nonparametric graph-based clustering method for discovering dissimilarity of the curve time series in space, which is justified with theoretical support to demonstrate how the model works under mild and easily verified conditions. We propose a very strict purity score that penalizes overestimation of clusters. Simulations show that our nonparametric graph-based clustering method is faster and more accurate than the parametric clustering method regardless of the size of data sets. We provide a Bayesian information criterion (BIC) to identify multiple change points and calculate a confidence interval for a mean response. By applying the CSAS model to the collected data, we can explain the differences between prevention and control policies in China and selected countries. 
\end{abstract}


\allowdisplaybreaks

\newpage
\spacingset{1.45} 

\section{Introduction}
\begin{figure*} [htb!]
		\includegraphics[width=1\textwidth]{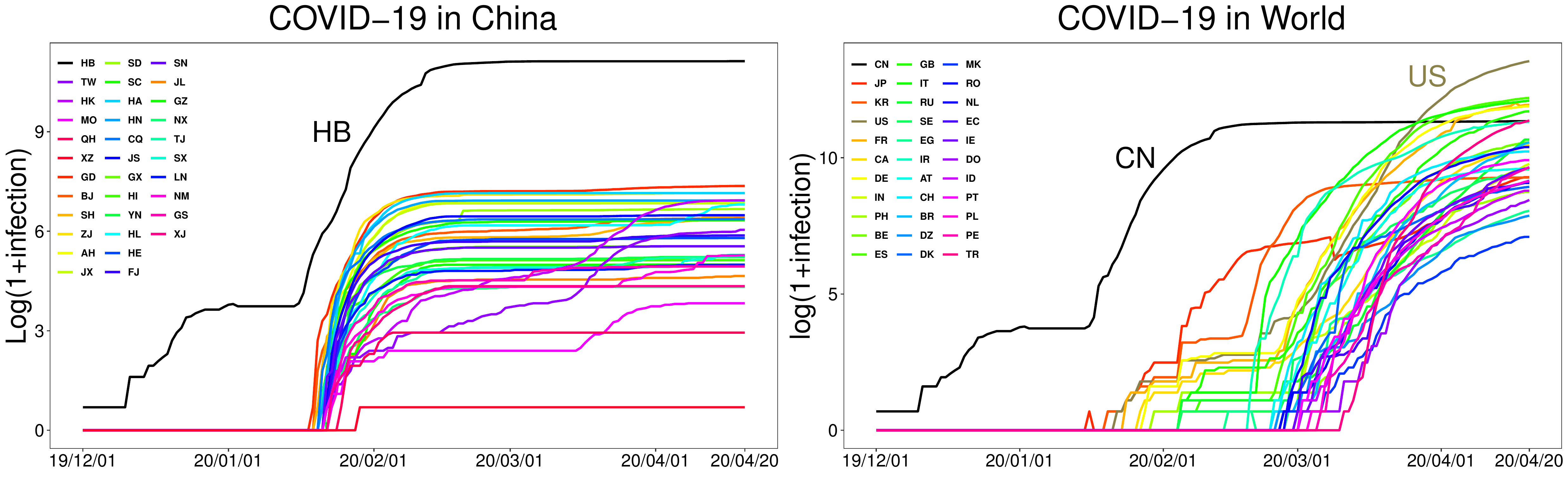}
		\caption{Plots of log-transformed  infectious counts from December 1, 2019 to April 20, 2020 in  China   and in 33 selected countries.
			  The data origin is from https://github.com/canghailan/Wuhan-2019-nCoV. The Alpha-2 codes applied here for China's provinces/regions and countries come from  https://www.iso.org/obp.
		}
		\label{fig1}
\end{figure*}

During  the COVID-19 outbreak, multiple complex factors resulted in the space-time pattern of  spread. Fig. \ref{fig1} shows the log-transformed infectious counts in each region in China, and in 33 selected countries at the end of April 20, 2020.

From Fig. \ref{fig1}, we can see two main characteristics of the spread: (i) the spread of COVID-19 has a  space-time characteristic determined by different intervention policies, incomplete information, geographical locations, transport, climate, and so on; (ii) along the time, the log-transformed infectious counts presented different sigmoid (stretched S-shaped) curves. This phenomenon often happens in  the life cycles of plants, animals, and viruses, which can rise and fall periodically. In each cycle, the sigmoid curve experiences three phases: slow rising, sharp rising, and slow falling.

Modeling the spread of COVID-19 in many regions over a long period of time is proven to be challenging. That is because many regions may not share the same spread pattern  and different regions  may exhibit various intervention policies that may cause instability in the models.
The model for each region may  have a large degree of noise, but a common cluster of all regions could have less noise by the law of large numbers. Thus, clustering is of importance to increase   model fit.  We may have to cluster all regions, even if the number of clusters is unknown. 
In addition, we should allow the model to incorporate unknown change points to further enhance the fitting performance.  
Ignoring the existence of change points may lead to poor model fitting and misleading model interpretation \citep{SWWW16}.
 Furthermore, it is often necessary to apply the model to  updated data from an extended period of time. In the extended period, old clusters need to be updated and new change points may occur. Models with incorporated clusters and change points should be flexible and adaptive to the new data.
  In the next step, we shall consider the nonlinear characteristics of the models.

Logarithmic transformation is often used for transforming  count data, which    includes zero values \citep{JWRH20} and  grows exponentially over time. The simplest formula for exponential growth of a function $y$ at the growth rate $r$, as time $t$ goes on,  is $y(t)=y(0)(1+r)^t$, which satisfies the linear differential equation $\frac{dy(t)}{dt}=\log(1+r)y(t)$. A nonlinear variation of this differential equation may lead $y(t)$ to a sigmoid function. For example, the solution of a nonlinear differential equation  $\frac{dy(t)}{dt}=\log(1+r)y(t)-y^2(t)$ is the logistic function \citep{M89,LS17}.  The exponential growth model has shown numerous applications in the modeling and controlling of complex systems. For example, the number of cells  in a culture will increase exponentially until an essential nutrient is exhausted. 
A virus, for example SARS or COVID-19,  has been found to spread exponentially \citep{KMBM20}. The speed of spread  slows down when an artificial immunization becomes available or intervention policies take effect.   Other applications of the exponential growth model can be found in Physics (e.g., radioactive decay), Economics (e.g., a country's gross domestic product), Finance (e.g., investments), Computer science (e.g., computing growth and  internet phenomena), and so on.

When systems have short-term memories and  become more complex, it is extremely difficult to find a differential equation to  describe the growth curve. In contrast, we may add some autoregressive terms in a regression function to adapt to the complex system.
\cite{KKMM17} shows that an autoregressive logistic model was more accurate than a logistic model  when it comes to predicting the behaviors of complex biological systems. The reason is that the added autoregressive terms, which behave like short-term memory,  can make an appropriate adjustment to better fit the complex system.
In the same spirit, we propose the 	clustering-segmented
autoregressive sigmoid  (CSAS) model with four key characteristics  including unknown clusters,  change points, stretched S-shaped curves, and  autoregressive terms. With the help of the CSAS model, we expect to understand how an outbreak is spreading in time and in space,  to understand how the spread is affected by epidemic control strategies, and to apply the model to updated data from an extended period of time.

To identify this CSAS model, we first identify unknown clusters.  There are many popular methods, such as 
 K-means \citep{WH79} (implemented in the \textit{R} function \textit{kmeans}), Expectation-Maximization clustering for Gaussian Mixture Models (GMM-EM) \citep{A95} (implemented in the \textit{R} package \textit{mclust}), Density-Based Spatial Clustering of Applications with Noise (DBSCAN) \citep{EKSX96} (implemented in the \textit{R} function \textit{fpc::dbscan}), and Hierarchical clustering \citep{ML14} (implemented in the \textit{R} function \textit{hclust}).
Except for GMM-EM, which can be considered to be parametric,   all other methods  need to predetermine the number of clusters or distance related parameters. To compare the dissimilarity of the curve time series, we need a nonparametric method that does not require predetermined parameters. Then, we can separate  different  regions from China and the selected 33 countries  into clusters that share common patterns, segment the curve time series, and provide  accurate fittings.

 Our contributions include the following: (1) we propose the CSAS model to help  understand how an outbreak is spreading in time and in space,  to understand how the spread is affected by epidemic control strategies, and to apply the model to updated data from an extended period of time;
 (2) we provide a nonparametric graph-based clustering method with theoretical support, which furthermore proposes a very strict purity score that   penalizes the overestimation of clusters.  
 Simulations show that our method is fast and efficient for different sizes of data sets; (3) we give  practical methods for segmentation and  provide a confidence interval estimation for mean response; (4) we analyze the COVID-19 data in   regions in China and selected countries, and explain the differences among the  epidemic prevention and control policies.

\section{Main results}

We assume the clustering-segmented autoregressive
	sigmoid (CSAS) model:
\begin{align}\label{M0}
Z_{i, t}=&\sum_{m=1}^{M_i}\bigg\{\beta^{(m)}_{1,i}+\beta^{(m)}_{2,i}\Phi(\beta^{(m)}_{3,i}+\beta^{(m)}_{4,i}t)  \nonumber\\
&  +\sum_{q=1}^p\beta^{(m)}_{q+4,i}Z_{i, t-q}+\varepsilon_{i, t}^{(m)}\bigg\}I(\tau_i^{(m-1)}<t\leq \tau_i^{(m)}),
\end{align}
 where $Z_{i, t}=\log(1+Y_{i, t})$; $Y_{i, t}$ is the number of confirmed cases for the $i$th ($1\leq i\leq N$) cluster and time $t\in [1, T]$;  $i=\delta(j)$ for $j$th region with $1\leq j\leq K$; $Z_{i, 1-q}=0$ for $q=1,\ldots,p$;  $I(A)$ is an indicator function taking 1 if $A$ is true, 0 otherwise;  $\tau_i^{(0)}=0$, $\tau_i^{(M_i)}=T$, $\tau_i^{(m)}$ for $M_i>1$ and $1\leq m\leq M_i-1$ are common change points for the $i$th cluster;  $\Phi(x)=\frac{1}{\sqrt{2\pi}}\int_{-\infty}^xe^{-u^2/2}du$ is a  cumulative distribution function (CDF) of the standard normal distribution representing the  sigmoid  curve;  
	$\beta^{(m)}_{1,i}$'s and $\beta^{(m)}_{2,i}$'s are stretch location and scalar parameters, respectively;  $\beta^{(m)}_{3,i}$'s and $\beta^{(m)}_{4,i}$'s
	are linear regression coefficients within the  sigmoid curves; $\beta^{(m)}_{q+4,i}$'s  are autoregressive regression coefficients;  $\varepsilon_{i, t}^{(m)}$'s are independent random errors with a mean of zero and constant variance of $(\sigma_i^{(m)})^2$. 

The  CSAS  model has four key characteristics: (1) it is implemented  with unknown $N$ different clusters among $K$ regions. Due to the epidemic mechanism, human mobility and control strategy, the spread of epidemics displays a spatial propagation. We will propose a nonparametric method to cluster the regional data by applying the characteristic of sigmoid curve. This method does not introduce any factors and hence can be considered   nonparametric;
(2) the multiple S-shaped curves are described by change points. The change points $\tau_i^{(m)}$ for $1\leq m\leq M_i-1$ are unknown and are related to the cluster ($i$).   This is because different intervention policy releases such as lockdown, maintaining social distance,  cancelling large events, closing schools, and so on, result in different segmented sigmoid curves among unknown clusters;
(3) the regression function is mainly determined by the stretched S-curve $\beta^{(m)}_{1,i}+\beta^{(m)}_{2,i}\Phi(\beta^{(m)}_{3,i}+\beta^{(m)}_{4,i}t)$  and it allows a slight adjustment through the autoregressive terms, $\sum_{q=1}^p\beta^{(m)}_{q+4,i}Z_{i, t-q}$ which can be considered as a short-term memory for the response variable;  and (4) after specifying both clusters and change points, we use the corresponding data sets to answer three questions.  How do we estimate the regression coefficients? Are those coefficients significantly different from zero?
How do we give a confidence interval for the mean response?

  We give the following five remarks for the logarithmic transformation, the CDF function $\Phi(x)$, and the random error in the CSAS model.

{\bf Remark 1.} $\log(1+x)$ transformation is often used for transforming  count data that  include zero values \citep{JWRH20}.   When $Y_{i, t}$ is much smaller or larger than 1 in magnitude, $\log(1+Y_{i, t})\approx Y_{i, t}$ or $\log(1+Y_{i, t})\approx \log Y_{i, t}$ can be used. This transformation  $\log(1+Y_{i, t})$ of $Y_{i, t}$, which may  grow exponentially over time, has two patterns, slow rises and slow falls, and hence can often be modeled by a stretched S-curve.

{\bf Remark 2.} The nonlinear function $\Phi(x)$ is used to describe the stretched S-shaped curve. Other similar functions may be considered. For example, if we apply the approximation of $\Phi(x)$ by
\citep{T63}, $\Phi(x)\approx\frac{1}{1+e^{-2\sqrt{2/\pi}x}}$ for all $x$, then we have
$$\beta^{(m)}_{1,i} +\beta^{(m)}_{2,i}\Phi(\beta^{(m)}_{3,i}+\beta^{(m)}_{4,i}t)\approx\beta^{(m)}_{1,i}+\frac{\beta^{(m)}_{2,i}}{1+e^{-2\sqrt{2/\pi}(\beta^{(m)}_{3,i}+\beta^{(m)}_{4,i}t)}},$$
which is an extended  logistic function of time $t$ and is commonly used in logistic regression.

{\bf Remark 3.}
Mathematical modelling may provide an understanding of spread mechanisms. The original mathematical model was proposed and solved by
Daniel Bernoulli in 1760; see \citep{DH2002}. Recent developments and applications are mainly focused on the susceptible-infectious-recovered (SIR) model and its variants.
The logistic function derived from a nonlinear differential equation may explain why we should apply the sigmoid curve to model the spread of disease \citep{M89, LS17,KMBM20}.

{\bf Remark 4.}    The model for each region may  have a large degree of noise, but a common cluster of all regions could have less noise because of the law of large numbers. So, we should consider an individual cluster in the CSAS model.  From Fig. \ref{fig7}  C (Cluster 3), it can be seen that the noise  is significantly smaller than that of  Fig. \ref{fig7}  A (Province NM) or B (Province TJ).
In addition, we should allow the model to incorporate unknown change points to further enhance the fitting performance.
Fig. \ref{fig8} F suggests that the  residuals from the CSAS model without change points exhibit a clear trend. 
In contrast, the variance of noises in each segment  should be constant; see Fig. \ref{fig8} A-C. 
Models with incorporated clusters and change points should be flexible and adaptive to the new data;  
 see the continued good performance of the CSAS model in the    extended two-month data in the ``Discussion and Conclusions'' section.

{\bf Remark 5.}
With both  autoregressive terms and  CDF function in the CSAS model, the variance  of random errors can be considered to be
constant across segments.
 In Fig. \ref{fig8} A, B and C,  the model residuals are well-behaved across segments. The residuals in
Fig. \ref{fig8} D (the autoregressive terms are removed)  and Fig. \ref{fig8}  E (the CDF function is removed as shown in the Long-Short-Term-Memory model in \citep{Y20}) suggest a time trend. 
This finding agrees with the fact that an autoregressive logistic model was more accurate than  a logistic model as shown in \citep{KKMM17}. 

\subsection{Clustering}
To find clusters of all $K$ regions, we consider the $T$ dimensional series $\{\bm{Z}_j, 1\leq j\leq K\}$, where its $t$th component is $Z_{j,t}$ for $1\leq t\leq T$, and define the Euclidean distance between $\bm{Z}_{j_1}$ and $\bm{Z}_{j_2}$ as follows:

\begin{equation}\label{dist}
d(\bm{Z}_{j_1}, \bm{Z}_{j_2})=\sqrt{\frac{1}{T}\sum_{t=1}^T(Z_{j_1, t}-Z_{j_2, t})^2}.
\end{equation}

We construct an approximate shortest Hamiltonian path (SHP) based on a heuristic Kruska algorithm (HKA), which was proposed by \citep{BMG14} for a two-sample test. This was successfully applied into change point detection in \citep{SWR17,SWR18}. 
The HKA first sorts all edges in order of increasing distance defined in \eqref{dist}. First and foremost, the edge with a minimum distance must be selected. Then  subsequent edges are  chosen one-by-one from  the remaining list of sorted edges according to the requirement of a path. If this current edge does not form a cycle with the previously selected edges, and every vertex connected by this current edge, or previously selected edges, has a degree not greater than 2, then this current edge must be selected.
The HKA terminates when $K-1$ edges have been chosen. 
The approximate SHP is formed by chosen $K-1$ edges  denoted as  $\mathcal{P}=(j_1,\ldots, j_K).$ The next step is to find clusters based on $\mathcal{P}.$ We define the edge set of  $\mathcal{P}$ as $\mathcal{E}(\mathcal{P})$, and consider a subset of $\mathcal{E}(\mathcal{P})$:
\begin{eqnarray}&\label{E}
\mathcal{E}^*(\mathcal{P}, \theta)=\left\{(j_s, j_{s+1})~\text{for}~s=1,\ldots,K-1\right.\\ \nonumber
&\left.\text{such that}~ (j_s, j_{s+1})\in\mathcal{E}(\mathcal{P})~\text{and}~ d(\bm{Z}_{j_s}, \bm{Z}_{j_{s+1}})\leq \theta\right\}.
\end{eqnarray}

We create a graph from  the edge set $\mathcal{E}^*(\mathcal{P}, \theta)$  and define the connected components of this graph as a set of clusters $\mathcal{A}=\{\mathcal{A}_\ell, 1\leq \ell\leq L\}$. We note that the R function \textit{components} in  the \textit{R} package \textit{igraph} \citep{CN06} can calculate the connected components given the edge set.

Suppose that there is a set of classes $\mathcal{C}=\{\mathcal{C}_i, 1\leq i\leq N\}$, where $\mathcal{C}_i=\{j | \delta(j)=i\}$.
We need to measure
how close the set of clusters $\mathcal{A}$ is to the predetermined set of classes $\mathcal{C}$.
Purity    \citep{MRS08} is a measure of this extent   defined as:
\begin{equation}\label{Purity}
S(\mathcal{A}, \mathcal{C})=\frac{1}{K}\sum_{\ell=1}^L\max_{1\leq i\leq N}|\mathcal{A}_\ell\cap\mathcal{C}_i|.
\end{equation}
In most cases, a bad clustering has a purity value close to 0 and a perfect clustering has a purity of 1. However, this measure may not give a realistic evaluation for overestimated clusters. For example,  a purity score of 1 could happen by putting $\mathcal{A}_\ell=\ell$, $L=K$ and $N=1$. In this case, one whole class is mis-clustered to $K$ separate clusters with a purity score of 1.

We propose a very strict purity score to penalize overestimated clusters:
\begin{equation}\label{SPurity}
S^*(\mathcal{A}, \mathcal{C})=\frac{1}{K}\sum_{\ell=1}^L\max_{1\leq i\leq N}|\mathcal{A}_\ell\cap\mathcal{C}_i|-\frac{|L-N|}{\max(L, N)}.
\end{equation}
Users may add additional weight on the second penalty term according to different requirements.
Based on this very strict purity evaluation, a very bad clustering would have a purity value close to -1, and a perfect clustering will still have a purity of 1. If $\mathcal{A}_\ell=\ell$, $L=K$ and $N=1$, then $S^*(\mathcal{A}, \mathcal{C})=1/L$, decreasing as $L$ increases.
Overestimated clusters may have a very strict purity score close to 0.  A natural question comes: does our clustering have a very strict purity score of 1? To answer this question, we make the following assumptions.

{\bf Assumption 1.} Let $\varepsilon_{i, t}$ be $Z_{i, t}-E(Z_{i, t})$. Assume that $\varepsilon_{i, t}$ is independent and identically distributed (i.i.d.) satisfying $E(\varepsilon^4_{i, t})<\infty$ for all $1\leq j\leq K$ and $1\leq t\leq T$.

{\bf Assumption 2.} There exists a $\eta(T)$, satisfying that $\eta^2(T)>2E(\varepsilon^2_{1, 1})$, $K<<\{\eta^2(T)-2E(\varepsilon^2_{1,1})\}^2T$ and
$\min_{j_1\neq j_2, \delta(j_1)\neq \delta(j_2)}
d(E(\bm{Z}_{j_1}), E(\bm{Z}_{j_2}))>2\eta(T).$

In Assumption 1, if $\varepsilon_{i, t}$ is dependent, then we require the upper bound of $$E\left|\sum_{t=1}^T(\varepsilon_{j_1,t}-\varepsilon_{j_2,t})^2-E(\varepsilon_{j_1,t}^2)-E(\varepsilon_{j_2,t}^2)\right|^2<<T\eta^2(T)/K.$$ In Assumption 2, we require $K$ to be quite small compared to $T$. Note that $d(E(\bm{Z}_{j_1}), E(\bm{Z}_{j_2}))$ is easy to evaluate because
$d(E(\bm{Z}_{j_1}), E(\bm{Z}_{j_2}))=\sqrt{\frac{1}{T}\sum_{t=1}^T\{E(Z_{j_1, t})-E(Z_{j_2, t})\}^2}.$ We have the following Theorem 1.

{\bf Theorem 1.} Suppose Assumptions 1-2 hold. Choose $\theta=\eta(T)$ as in \eqref{E}. As $T\rightarrow\infty$, we have $P\{S^*(\mathcal{A}, \mathcal{C})=1\}\rightarrow1.$

Proof of Theorem 1. We first prove that $P\{\max_{j_1\neq j_2, \delta(j_1)=\delta(j_2)}\sqrt{\frac{1}{T}\sum_{t=1}^T(Z_{j_1, t}-Z_{j_2, t})^2}>\eta(T)\}\rightarrow0.$ Because $\delta(j_1)=\delta(j_2)$, $Z_{j_1, t}-Z_{j_2, t}=\varepsilon_{j_1, t}-\varepsilon_{j_2, t}$. Then we have
\begin{align}\label{P1}
&P\left\{\max_{j_1\neq j_2, \delta(j_1)=\delta(j_2)}\sqrt{\frac{1}{T}\sum_{t=1}^T(Z_{j_1, t}-Z_{j_2, t})^2}>\eta(T)\right\}\nonumber\\
&\leq \sum_{j_1\neq j_2}P\left\{\sum_{t=1}^T(\varepsilon_{j_1, t}-\varepsilon_{j_2, t})^2>\eta^2(T)T\right\}\nonumber\\
&=\sum_{j_1\neq j_2}P\Bigg[\sum_{t=1}^T(\varepsilon_{j_1, t}-\varepsilon_{j_2, t})^2-E\{(\varepsilon_{j_1, t}-\varepsilon_{j_2, t})^2\} \nonumber\\
&>\{\eta^2(T)-2E(\varepsilon^2_{1,1})\}T\Bigg]\nonumber\\
&\leq\sum_{j_1\neq j_2}P\Bigg[\left|\sum_{t=1}^T(\varepsilon_{j_1, t}-\varepsilon_{j_2, t})^2-E\{(\varepsilon_{j_1, t}-\varepsilon_{j_2, t})^2\}\right|^2\nonumber\\
&>\{\eta^2(T)-2E(\varepsilon^2_{1,1})\}^2T^2\Bigg] \nonumber\\
&\leq \frac{cK}{\{\eta^2(T)-2E(\varepsilon^2_{1,1})\}^2T},
\end{align}
where $c$ is  a constant not related to either $K$ or $T$. By Assumption 2, this upper bound converges to zero.
Next, we prove that  $P\{\min_{j_1\neq j_2, \delta(j_1)\neq\delta(j_2)}\sqrt{\frac{1}{T}\sum_{t=1}^T(Z_{j_1, t}-Z_{j_2, t})^2}\leq\eta(T)\}\rightarrow0.$ By the Minkowski inequality and  Assumption 2,

\begin{align*}
&P\left\{\min_{j_1\neq j_2, \delta(j_1)\neq\delta(j_2)}\sqrt{\frac{1}{T}\sum_{t=1}^T(Z_{j_1, t}-Z_{j_2, t})^2}\leq\eta(T)\right\}\nonumber\\
&\leq P\Bigg[\min_{j_1\neq j_2, \delta(j_1)\neq\delta(j_2)}\sqrt{\frac{1}{T}\sum_{t=1}^T\{E(Z_{j_1, t})-E(Z_{j_2, t})\}^2}\nonumber\\
&-\sqrt{\frac{1}{T}\sum_{t=1}^T\{\varepsilon_{j_1, t})-\varepsilon_{j_2, t})\}^2}\leq\eta(T)\Bigg]\nonumber\\
&\leq P\Bigg[\min_{j_1\neq j_2, \delta(j_1)\neq\delta(j_2)}2\eta(T)-\sqrt{\frac{1}{T}\sum_{t=1}^T\{\varepsilon_{j_1, t})-\varepsilon_{j_2, t})\}^2}\leq\eta(T)\Bigg]\nonumber\\
&= P\Bigg[\max_{j_1\neq j_2, \delta(j_1)\neq\delta(j_2)}\sqrt{\frac{1}{T}\sum_{t=1}^T\{\varepsilon_{j_1, t})-\varepsilon_{j_2, t})\}^2}\geq\eta(T)\Bigg], \nonumber
\end{align*}

which converges to zero by \eqref{P1}.

By the HKA, for any $\mathcal{A}_\ell$, there exists $\mathcal{C}_i$ such that $\mathcal{C}_i=\mathcal{A}_\ell$ in probability, which implies that $\max_{1\leq i\leq N}|\mathcal{A}_\ell\cap\mathcal{C}_i|=|\mathcal{A}_\ell|$ and $L=N$ hold in probability. So, $P(S^*(\mathcal{A}, \mathcal{C})=1)$ converges to 1 as $T\rightarrow\infty$. The proof of Theorem 1 is finished.

To apply Theorem 1, we need to set the right value for $\theta$.   In real problems, $\theta$ could be unknown. We shall propose a
data driven method to select the threshold value of $\theta$.  A naive choice of $\theta$ based on outlier detection is
\begin{align}\label{thre}
&\hat\theta=\text{median}_{s=1,\ldots,K-1}(x_s)\nonumber \\
&+2.5\left(1.483\times\text{median}_{s=1,\ldots,K-1}|x_s-\text{median}_{s=1,\ldots,K-1}(x_s)|\right),
\end{align}
where $x_s=d(\bm{Z}_{j_s}, \bm{Z}_{j_{s+1}})$ for $s=1,\ldots,
K-1$,  $\text{median}_{s=1,\ldots,K-1}(x_s)$ and $1.483\times\text{median}_{s=1,\ldots,K-1}|x_s-\text{median}_{s=1,\ldots,K-1}(x_s)|$ are  robust estimates of mean and  standard deviation of $\{x_s, s=1,\ldots,K-1\}$, respectively, and 2.5 is the cutoff value.
It works well for relatively small $N$ to $K$. The large values in the series of $\{x_s, s=1,\ldots,K-1\}$  would not affect the threshold value $\hat\theta$ and hence they could be successfully removed. However, if the distribution of $x_s$'s, with the exception of outliers, is  a mixture of two or more probability distributions which commonly occurs in multiple clusters, then $\hat\theta$ may not be consistent to $\theta$.  Therefore, we propose  Algorithm 1 based on Bayesian information criterion (BIC) to choose $\theta$.

\subsection{Segmentation}
Denote a set of change points as $C_i=\{\tau_i^{(1)}, \cdots, \tau_i^{(M_i-1)} \}$, where $i=1,\ldots,N$ and  $M_i-1$ is the number of change points.
Since $M_i$ is unknown in practice, we  would need to estimate the change points. Consider the segment  $[t_-, t^+]$ and define two residual sums of squares
\begin{align}
&S_{i,0}(t_-, t^+)
=\min_{\bm\beta}\sum_{t=t_-}^{t^+}\left\{{Z}_{i,t}-f(t; \bm\beta)\right\}^2,\label{ST1}\\
&S_{i,1}(t_-, t_0, t^+)
=\min_{\bm\beta}\sum_{t=t_-}^{t_0}\left\{{Z}_{i,t}-f(t; \bm\beta)\right\}^2\nonumber\\
&+\min_{\bm\beta}\sum_{t=t_0+1}^{t^+}\left\{{Z}_{i,t}-f(t; \bm\beta)\right\}^2\label{ST2},
\end{align}
 where  $1\leq t_-<t_0< t^+\leq T$ and   $f(t; \bm\beta)=\beta_1+\beta_2\Phi(\beta_3+\beta_4t)+\beta_5Z_{i, t-1}+\beta_6Z_{i, t-2}$. Here, we consider two autoregressive terms.   Then, the estimated change point  is denoted as $\hat t_i(t_-, t^+)$:
\begin{equation}\label{fit}
\hat t_{i, t_-, t^+}=\arg\min_{t_-+\Delta/2<t_0<t^+-\Delta/2}S_{i,1}(t_-, t_0, t^+).
\end{equation}
where $\Delta$ is  the minimum distance between two adjacent change points and $t^+-t_->\Delta$.

In light of  \cite{BP03}, we apply the  BIC method for model comparison. Define  
\begin{equation}\label{BIC}
\text{BIC}_{i,\nu} (t_-, t^+)=(t^+-t_-+1)\log\{{\hat\sigma_{i,\nu}}^2\}+6(\nu+1)\log(t^+-t_-)
\end{equation}
where $\nu=0$ or $1$, $6(\nu+1)$ is the number of parameters  and ${\hat\sigma_{i,0}}^2=(t^+-t_-+1)^{-1}S_{i,0}(t_-, t^+)$ and  ${\hat\sigma_{i,1}}^2=(t^+-t_-+1)^{-1}S_{i,1}(t_-, \hat t_{i, t_-, t^+}, t^+)$.
 Combined with the Iterated Cumulative Sums of Squares Algorithm (ICSS) \citep{IT94}, we propose Algorithm 2 to estimate multiple change points.

In Algorithm 2, there are two main steps that include finding candidate change points and refining  them. We set the minimum distance between two adjacent change points, $\Delta$, to be  $10$ for real data analysis.

 \subsection{Fitting} First, we use the well-known \textit{nls} function in the  \textit{R}   package  \textit{stats} \citep{R20}  to find the minimum value as shown in \eqref{ST1} and give  t tests on regression coefficients,
 where initial values of parameters are given by grid search. Second,  we give a confidence interval of regression function, denoted as $g_{i,t}(\bm\beta)=E(Z_{i,t}|Z_{i,t-1},Z_{i,t-2})$ for $t\in[t_-, t^+]$, by the delta method as follows.  By first-order Taylor expansion at the solution $\hat{\bm\beta}$, we have
 \begin{equation*}
g_{i,t}(\bm\beta)\approx g_{i,t}(\hat{\bm\beta})+\nabla{g_{i,t}(\hat{\bm\beta})}(\bm\beta-\hat{\bm\beta}).
 \end{equation*}
 The approximate $(1-\alpha)100\%$ confidence interval for  $g_{i,t}(\bm\beta)$ is
 \begin{align*}
g_{i,t}(\hat{\bm\beta})\pm t^*_{\alpha/2}(t^+-t_--6)\sqrt{\nabla{g_{i,t}(\hat{\bm\beta})}\top\text{Var}(\hat{\bm\beta})\nabla{g_{i,t}(\hat{\bm\beta})}},
 \end{align*}
 where $t^*_{\alpha/2}(t^+-t_--6)$ is the
  $\alpha/2$ lower quantile of a t distribution with degrees of freedom $t^+-t_--6$ and $\text{Var}(\hat{\bm\beta})$ can be estimated by the \textit{nls} function in \textit{R}. Here, we consider $\alpha=0.05$.

\section{Simulations}
We consider three classes $N=3$. Let $\mathcal{C}_1, \mathcal{C}_2,$ and $\mathcal{C}_3$ be randomly seperated classes with $\cup_{i=1}^3\mathcal{C}_i=\{1,\ldots,K\}$. Denote the number of elements in $i$th class as $n_i$ with $\sum_{i=1}^3n_i=K$.
We produce $Z_{i,t}$   from the following model
\begin{align}\label{M1}
Z_{i, t}&=\sum_{m=1}^{M_i}\left\{\beta^{(m)}_{1,i}+\beta^{(m)}_{2,i}\Phi(\beta^{(m)}_{3,i}+\beta^{(m)}_{4,i}t)\right\}I(\tau_i^{(m-1)}<t\leq \tau_i^{(m)})\nonumber\\
&+\varepsilon_{t},
\end{align}
where $i=1,2,3$, $t=1,\ldots,T$ and $\varepsilon_{t}$'s are independent Normal errors with mean zero and variance $\sigma^2$.

For the first class, let $M_1=1$, $\beta^{(1)}_{1,1}=0$, $\beta^{(1)}_{2,1}=10$, $\beta^{(1)}_{3,1}=-4$, and $\beta^{(1)}_{4,1}=-0.05$.
For the second class, let $M_2=2$, $\beta^{(1)}_{\ell,2}=0$ for $\ell=1,\ldots,4$, $\tau_2^{(1)}=T/3$, $\beta^{(2)}_{1,2}=0$,
$\beta^{(2)}_{2,2}=20$,  $\beta^{(2)}_{3,2}=-3$, and  $\beta^{(2)}_{4,2}=0.03$.
For the third class, let $M_3=2$, $\beta^{(1)}_{\ell,3}=0$ for $\ell=1,\ldots,64$, $\tau_3^{(1)}=2T/3$, $\beta^{(2)}_{1,3}=0$,
$\beta^{(2)}_{2,3}=5$,  $\beta^{(2)}_{3,3}=-2$, and  $\beta^{(2)}_{4,3}=0.07$.

\begin{figure*} [ht]
 		\includegraphics[width=1\textwidth]{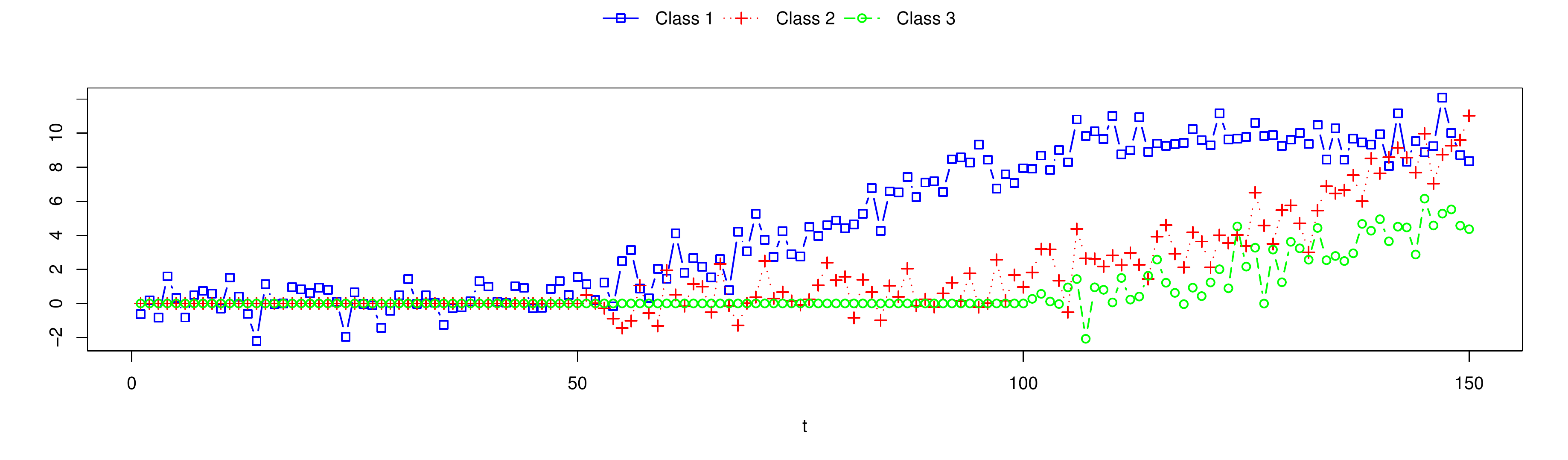}
 		\caption{Plots of S-curves for three classes.
 		}
\label{fig2}
 \end{figure*}

\begin{figure*} [ht]
 		\includegraphics[width=1\textwidth]{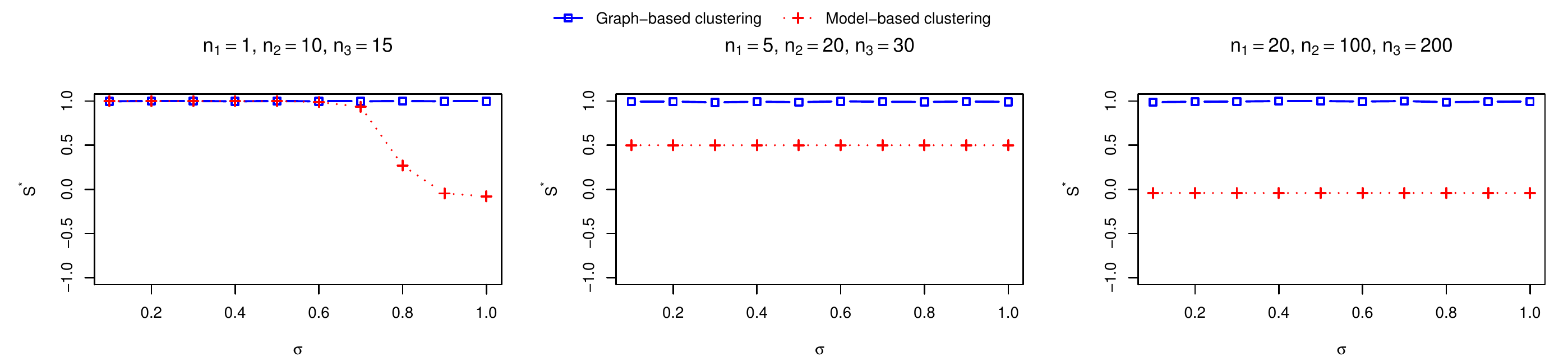}
 		\caption{Comparisons of graph-based clustering method and model-based clustering method.
 		}
\label{fig3}
 \end{figure*}

Fig. \ref{fig2} plots  different S-curves of $Z_{i,t}$ for $i=1,2,3$ and $T=150$. Next, we  compare our graph-based clustering method as shown in Algorithm 1 with the model-based clustering method \citep{FR02} implemented in the R function \textit{Mclust} \citep{FR20}. Fig. \ref{fig3} shows the averaged strict purity score as in \eqref{SPurity} for estimated clusters based on these two methods for  $\sigma=0.1, 0.2, \ldots,1$, different $n_i$'s and 100 replications.

It can be seen from Fig. \ref{fig3} that our method is very accurate for different $\sigma$ and sizes of classes because its very strict purity scores are close to 1. This agrees with the conclusion in Theorem 1.  In contrast, the performance of the model-based clustering method is affected by large $\sigma$ and large sizes of classes. Specially, when $n_1=20$, $n_2=100$ and $n_3=200$, the computation is significantly slower compared with our method. The comparison of computing time is not presented here.

\section{Real data analysis}
We continue to use the data set of log-transformed infection counts from December 1, 2019 to April 20, 2020 from Chinese provinces/regions and the  33 countries, and  present the clustering and S-shaped fitting with change points. Here, two autoregressive components ($p=2$) in \eqref{M0}  are suggested.
\subsection{Clustering}
\begin{figure*} [ht]
 		\includegraphics[width=1\textwidth]{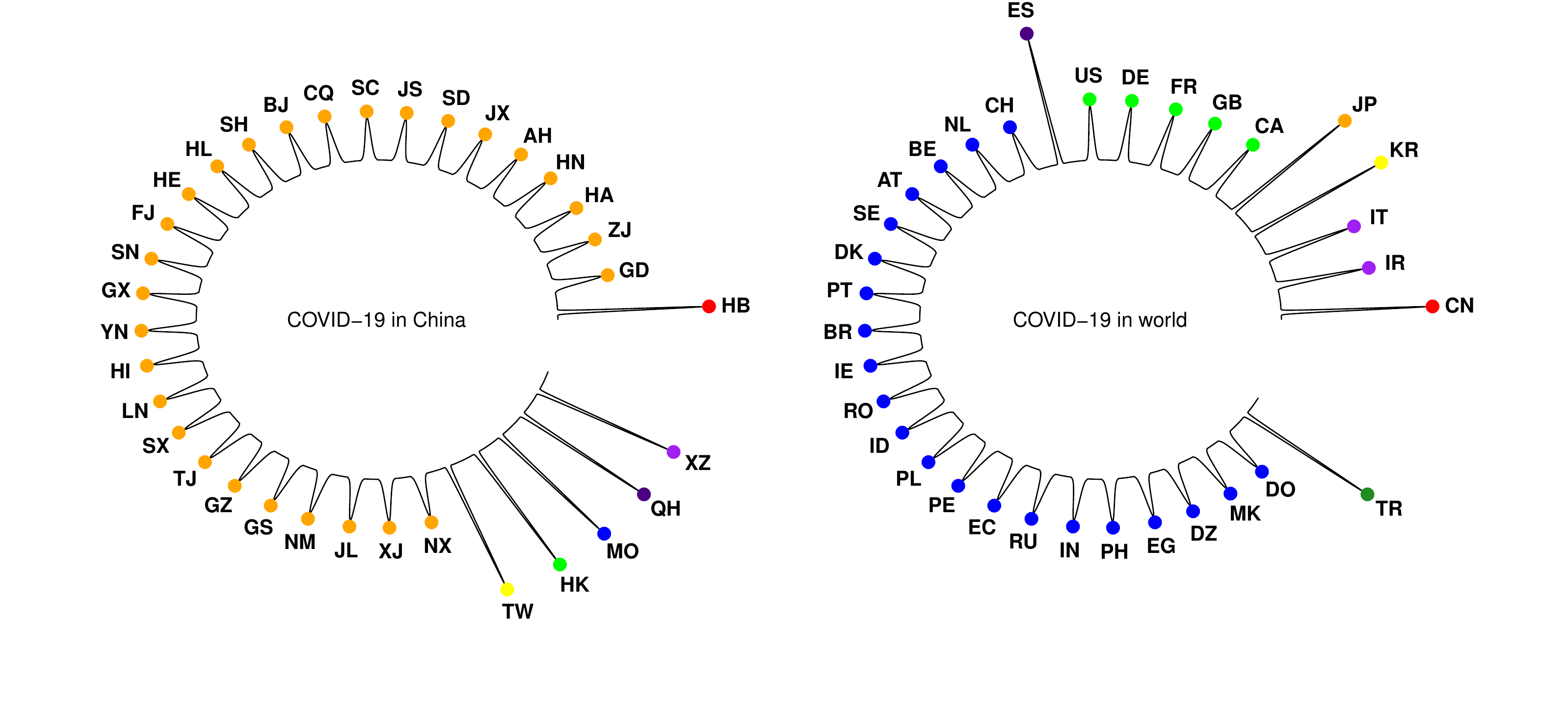}
 		\caption{Plots of clusters in China and in 33 selected countries  based on the log-transformed infection counts.}
\label{fig4}
 \end{figure*}

Based on the graph-based clustering Algorithm, the clusters of COVID-19 in China and the rest of the world are presented in Fig. \ref{fig4}, where the optimal path is presented as a cycle with vertexes representing clusters in different colors and overextended curves. This way of  presentation  is to transmit three aspects of information: (i) this analysis is for   virus data, therefore, we should use the cycle and the sharp nodes to describe the structure of the virus; (ii) the optimal graph is a path connecting all nodes where nodes can be provinces/regions in China or countries in the world; and (iii) readers can quickly find the different clusters and where to separate them from the path.

From Fig. \ref{fig4}, we observe the following.

(1) As shown in COVID-19 cases in China, the 34 provinces/regions are clustered into 7 categories. Specifically, Hubei (HB), Xizang (XZ), Qinghai (QH), Macao  (MO), Hong Kong (HK), and Taiwan (TW) are individually clustered into separate categories, and the remaining provinces/regions are all clustered into one category. This clustering result can be  explained by the differences in epidemic   control strategies among the provinces/regions: HB is the center of the COVID-19 breakout, with a large number of infection cases; underpopulated XZ and QH are both  located on the Qinghai Tibet Plateau, with only  a few infection cases; MO, HK, and TW are of self-governance: meaning their epidemic control strategies are different from all other regions in China. The model-based clustering method \citep{FR02} suggests both HK and TW are to be in one cluster, which may not be correct.

(2) As shown in COVID-19 cases in the world, the 33 selected countries are clustered into 8 categories. Specifically, China (CN), Korea (KR), Japan (JP), Spain (ES), and Turkey (TR) are individually clustered into separate categories; Italy (IT) and Iran (IR) are clustered into one category; the United States of America (US), Germany (DE), France (FR), the United Kingdom of Great Britain (UK), Northern Ireland (GB), and Canada (CA) are clustered into one category; and the remaining countries are all clustered into  one category. This clustering result is partly based on the timing of COVID-19 outbreaks in those countries. For example, the first large-scale outbreak was in CN, followed by KR and JP. After that,   infections in IR and IT experienced rapid growth, followed by the outbreaks in European countries and the US. Finally, the epidemic spread worldwide. In addition, the clustering is also based on the epidemic control strategies  in each country. For example, in KR and JP, even while the epidemic broke out around the same time, the two countries had taken different strategies: JP adopted a ``defensive strategy''' to ensure the health care system operated normally as usual, while KR used an ``aggressive attack strategy''  to comprehensively detect infections.

\subsection{Segmentation and fitting}
Based on the BIC-based ICSS Algorithm, we segment the curve time series and present the segmented  fittings and confidence interval estimation  for the  log-transformed infection counts $Z_{i,t} (1 \leq i \leq N)$ of each cluster in China and the rest of the world; see  Fig. \ref{fig5} and \ref{fig6}, respectively. 

We can obtain that all sigmoid curves share the form of multiple stages and multiple change points, with the exception of Cluster 7 (XZ) in China, with only one infection; the  calculated change points of each cluster can still be  explained by the differences in epidemic control strategies. See the  details below.

(1) As shown in Fig. \ref{fig5} A and Fig. \ref{fig6} A, the sigmoid curves and change points are almost the same because HB province was the center of the COVID-19 outbreak in CN. In Fig. \ref{fig6} A in CN, the first segment (19/12/01 to 19/12/13) was the germination period of the outbreak. In the second segment (19/12/13 to 20/01/16), COVID-19 seemed to have been controlled in CN. However, because many COVID-19 cases had not been found due to varied epidemic control strategies in the previous two stages, COVID-19 broke out in the  third segment (20/01/16 to 20/01/26) and  fourth segment (20/01/26 to 20/02/11) in CN. This coincided with  Chunyun (the annual massive movement of people during Chinese Lunar New Year), which particularly accelerated the outbreak. Finally, in the last two segments (20/02/11 to 20/02/27 and 20/02/27 to 20/04/20),  COVID-19 was controlled and stabilized once the CN government implemented very  strict epidemic control strategies, such as traffic control and home quarantine.

(2) As shown in Fig. \ref{fig5} C, the sigmoid curves in HK and TW seem similar because they were both strongly affected by  COVID-19 cases from mainland China. However, we find that the change points of COVID-19  in TW  are about a week delayed compared to those in HK after  COVID-19 started to break out in both regions. This is because TW responded in a timely manner to the  COVID-19 outbreak and controlled it  more quickly and effectively than HK, while the implementation of  epidemic control strategies in HK lagged behind.

(3) As shown in Fig. \ref{fig6}, the number of new cases in China had tentatively stabilized since the last change point, 20/02/27, which was delayed by about one week in other clusters. In Fig. \ref{fig6}  A and B, the infections in CN and KR are mostly stable, but the epidemic situations in other countries have not been controlled effectively. Take the fifth cluster (Fig. \ref{fig6}  C) as an illustration, considering that this cluster had the fastest growth. The four segments can be explained as follows: (i) the infections in the first segment were mainly from oversea imports; (ii) in the second segment, COVID-19 seemed to have been controlled; (iii) COVID-19 broke out because of many unfound COVID-19 cases in previous segments; and (iv) in the last segment, COVID-19 began to come under control as governments  declared states of emergency and started implementing strict measures to control the spread of the virus.

(4) As shown in Fig. \ref{fig6} B, confidence intervals for KR tended to be quite narrow in width when the number of new cases had tentatively stabilized, resulting in more precise estimates of mean response, whereas confidence intervals for JP tended to be wide since JP had adopted a ``defensive strategy''. In most of cases, confidence intervals produced precise results.
  
\section{Discussion and Conclusions}

A clustering-segmented autoregressive sigmoid model is developed to explore the space-time pattern of the log-transformed infectious count by the end of April 20, 2020. It performed well when it was applied to  COVID-19 cases in both China and the 33 countries, and thus provides an efficient statistical model of COVID-19 spread to help fight against the  virus. Currently, the infections in China are mostly stable, and the graph-based clustering algorithm is robust to the clusters from the 34 provinces/regions in China.  When COVID-19 began to come under control, the clustering of the disease globally will become increasingly stable.

\begin{figure*} [ht]
  		\includegraphics[width=1\textwidth]{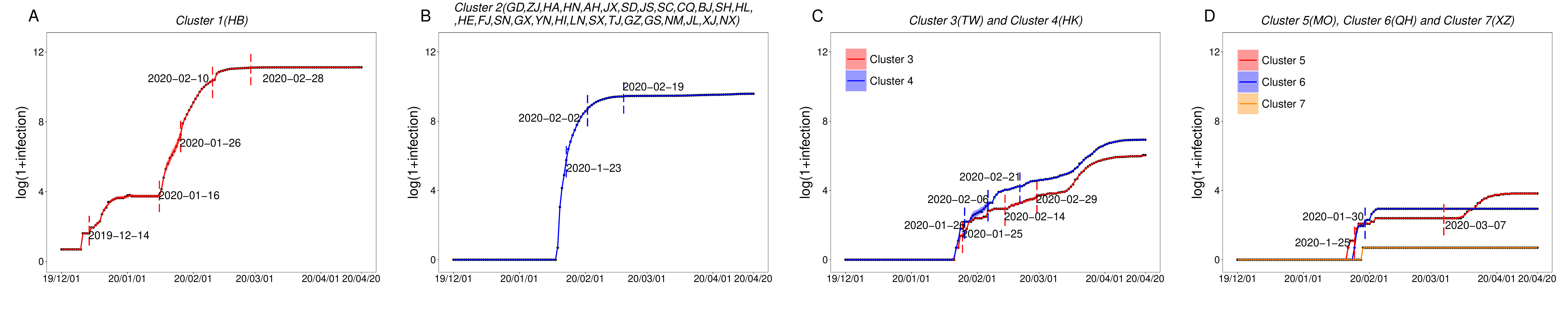}
  		\caption{Plots of segmentations and fittings of provinces/regions in China based on the log-transformed infection counts.
  		}
\label{fig5}
  \end{figure*}

\begin{figure*} [ht]
   		\includegraphics[width=1\textwidth]{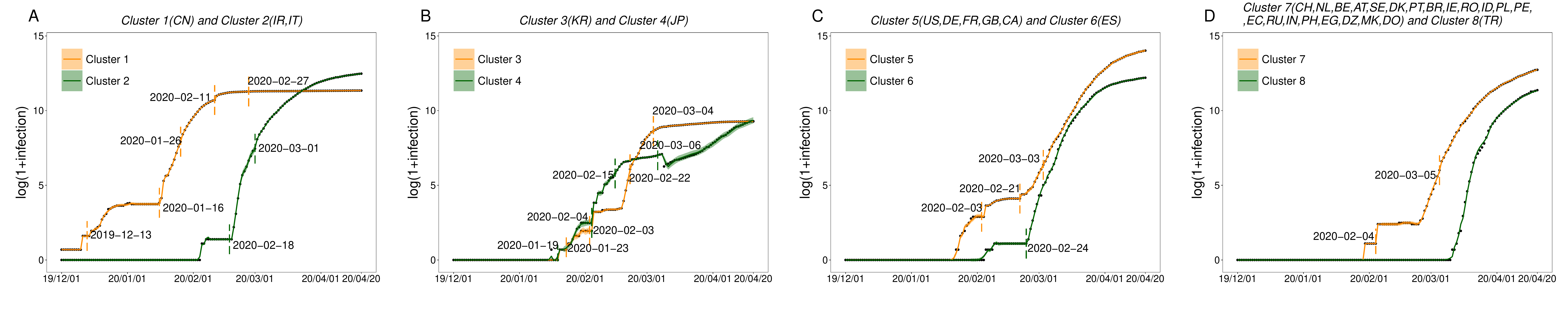}
   		\caption{Plots of segmentations and fittings of 33 selected countries in the world based on the log-transformed infection counts.
   		}
   		\label{fig6}
   \end{figure*}

\begin{figure*} [ht]
   		\includegraphics[width=1\textwidth]{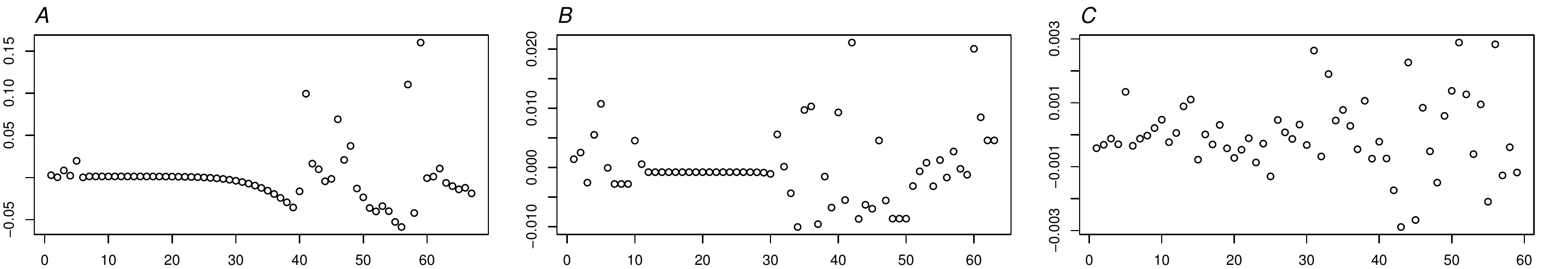}
   		\caption{ Plots of residuals of the last segment from the CSAS model for two separate provinces in China, NM (A) and TJ (B), and their common cluster 2 (C).   		}
\label{fig7}
   \end{figure*}

\begin{figure*} [ht]
   		\includegraphics[width=1\textwidth]{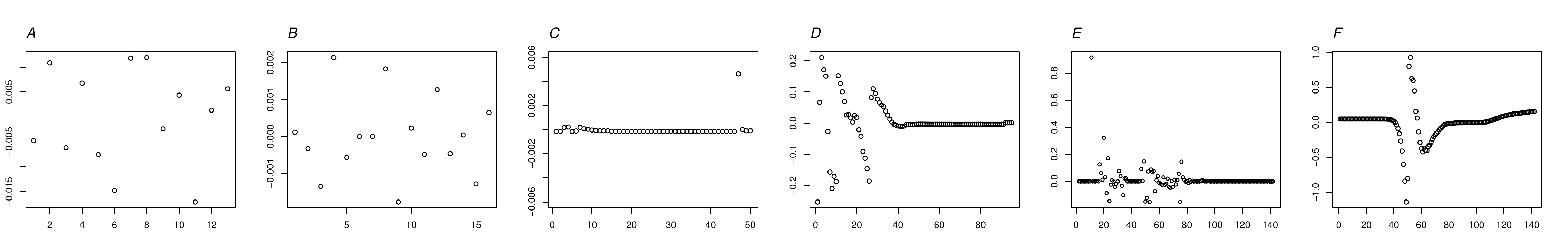}
   		\caption{  Plots of residuals for HB in China for the fourth segment (A), the fifth segment (B), and the last segment (C) from the CSAS model, for the last segment from the CSAS model without autoregressive terms (D), for the whole period from the autoregressive model of order 2 after taking the first difference (E), and for the whole period from the CSAS model without change points (F).   		}
   \label{fig8}
   \end{figure*}

\begin{figure*} [ht]
  	\includegraphics[width=1\textwidth]{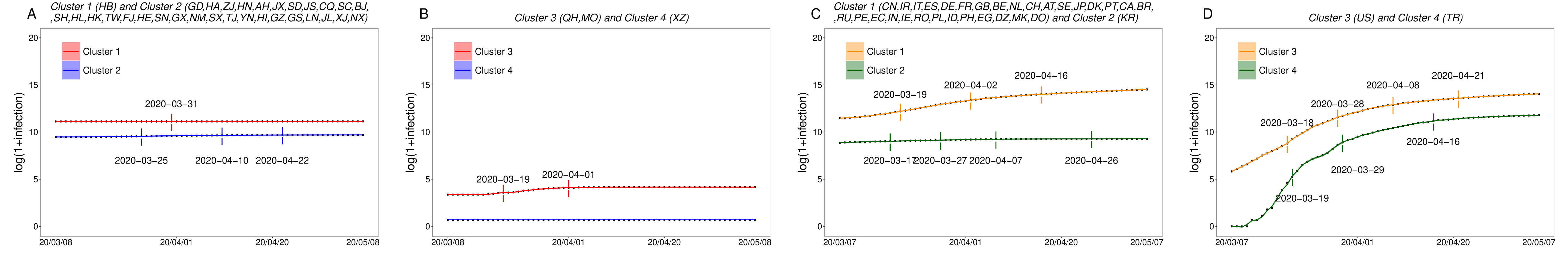}
  	\caption{Plots of segmentations and fittings of each cluster in China (A-B) and in 33 selected countries   (C-D) based on the log-transformed infection counts during the two-month extended period.}
  	\label{fig9}
  \end{figure*}

In fact, the CSAS model can adapt to an extended period of time when clusters have been updated and new change points have been identified. To do so, we use the last   change points in time, 20/03/07 obtained from Fig.  \ref{fig5} or 2020/03/08 obtained from Fig.  \ref{fig6}, as the start of the extended period at two-month intervals, from 20/03/07 to 20/05/07 or 20/03/08 to 20/05/08. In Fig  \ref{fig9}, we show segmentations and fittings for log-transformed infection counts of each cluster in both China and the 33 countries during this extended period. We can see that the fittings continue to work well. We provide an \textit{R} package, \textit{GraphCpClust}, which can be 
accessed from https://github.com/Meiqian-Chen/GraphCpClust. From this \textit{R} package, users can obtain the same  results presented in this paper and can model  data for another extended period of time. In addition, the data and code for another two papers \citep{SWR17,SWR18} are included in this \textit{R} package.

Regarding the dataset used in this article, Wuhan-2019-nCoV, we make the following additional remarks: 1. Back in early March 2020, there were very few datasets on COVID-19, and especially few datasets containing timely epidemic data from each Chinese province. This dataset, Wuhan-2019-nCoV, collects national outbreak reports from WHO, as well as daily outbreak reports from provincial health and family planning commissions in China; 2. The Wuhan-2019-nCoV dataset is very timely updated and has been included in the ``Open Source Wuhan'' data resource. Therefore, we believe that the data quality of the Wuhan-2019-nCoV dataset is trustworthy.  
There are now more and more COVID-19 data resources available, such as WHO data (https://covid19.who.int/) and Our World in Data (https://ourworldindata.org/covid-data-switch-jhu). For these two datasets, we find that our model still works very well. Please see   this webpage, http://graph-clustering-system.com/, for the three data analyses described above.

\newpage

 \begin{algorithm}[t]
 	\SetAlgoLined
 	\KwResult{Output the optimal clusters $\mathcal{A}$.}
 	{\bf Notations:} $x(\theta)=\{x_s | x_s\leq \theta, s=1,\ldots, K-1\}$ where $x_s$ is defined in \eqref{thre}\; $\theta_{(s)}$  is the $s$'th largest element in set $\{x_s, s=1,\ldots, K-1\}$\;
 	$\hat\sigma^2(\theta)$ is the sample variance of $x(\theta)$\;
 	$\text{BIC}(\theta, \mathcal{A})=(K-1)\log(\hat\sigma^2(\theta))+2L(\mathcal{A})\log(K-1)$, where $L(\mathcal{A})$ is the number of clusters in $\mathcal{A}$ \;
 	\textbf{Initialize}: Let $i=1$,  $L=1$, and $\mathcal{A}=\{\mathcal{A}_1\}$ where $\mathcal{A}_1=\{1,\ldots K\}$\;
 	
 	\For{$s = 2;\ s < K-1;\ s = s + 1$}{
 		Let $\theta$ be $\theta_{(s)}$ and  calculate  the clusters based on  $\mathcal{E}^*(\mathcal{P}, \theta)$ in \eqref{E} denoted as
 		$\mathcal{A}_{\text{temp}}$\;
 		\eIf{$\textup{BIC}(\theta, \mathcal{A}_{\text{temp}})<\textup{BIC}(\theta_{(s-1)}, \mathcal{A})$}{
 			$\mathcal{A}=\mathcal{A}_{\text{temp}}$\;
 		}  {break\;}

 	}
 	\caption{Graph-based clustering Algorithm}
 \end{algorithm}

\begin{algorithm}[t]
 	\caption{BIC-based ICSS Algorithm}
 	\SetAlgoLined
 	\KwResult{Output the estimated change points $\hat C_i$.}
  	{\bf Notations:} $\hat C_{i, (s)}$ and $|\hat C_i|$ are the $s$th smallest element and number of elements of set $\hat C_i$, respectively\;
 	{\bf Initialization:} Let $t_-=1$,  $t_+=T$, and $\hat C_i=\{0, T\}$\;
 	\While{$t^+-t_->\Delta$}{
 	$t_{\text{first}}\gets t^+$; $t_{\text{last}}\gets t_-$\;
 	
 	\While{$\textup{BIC}_{i,0} (t_-, t_{\textup{first}})\geq \textup{BIC}_{i,1} (t_-, t_{\textup{first}})$}{
   			$t_{\text{first}}\gets\hat t_{i, t_-, t_{\text{first}}}$;}
\While{$\textup{BIC}_{i,0} (t_{\textup{last}}, t^+)\geq\textup{BIC}_{i,1} (t_{\textup{last}}, t^+)$}{
   			$t_{\text{last}}\gets\hat t_{i, t_{\text{last}}, t^+}$\;}
   			
   			\eIf{$t_{\textup{first}}=t_{\textup{last}}$}{$\hat C_i\gets\hat C_i\cup\{t_{\textup{first}}\}$; break\;}
 				 {$\hat C_i\gets\hat C_i\cup\{t_{\text{first}}, t_{\text{last}}\}$; $t_-\gets t_{\text{first}}$; $t^+\gets t_{\text{last}}$\;}
 	}
 		\For{$s=2;\ j < |\hat C_i|;\ s = s + 1$}{
 			\If{$\textup{BIC}_{i,0} (\hat C_{i, (s-1)}+1, \hat C_{i, (s+1)})\leq \textup{BIC}_{i,1}(\hat C_{i, (s-1)}+1, \hat C_{i, (s+1)})$}{
 				$\hat C_i\gets\hat C_i\setminus \{\hat C_{i, (s)}\}$\;}
 				}
 	 $\hat C_i\gets\hat C_i\setminus\{0, T\}$\;
 	
\end{algorithm}

\end{document}